# Pulsars and Millisecond Pulsars III: Tracing Compact Object Dynamics in Globular Clusters with NBODY6++GPU

M. Rah[1,2]*, R. Spurzem[2,3,4], F. F. Dotti[5,6,3,7], A. Mickaelian[1]

[1] NAS RA V.Ambartsumian Byurakan Astrophysical Observatory (BAO), Byurakan, Armenia

[2] The Silk Road Project at the National Astronomical Observatory, Chinese Academy of Sciences, China

[3] Astronomisches Rechen-Inst., ZAH, Univ. of Heidelberg, Germany

[4] Kavli Institute for Astronomy and Astrophysics, Peking University, China

[5] Department of Physics, New York University Abu Dhabi, PO Box 129188 Abu Dhabi, UAE

[6] Center for Astrophysics and Space Science (CASS), New York University Abu Dhabi, PO Box 129188, Abu Dhabi, UAE

[7] Dipartimento di Fisica, Sapienza, Università di Roma, P.le Aldo Moro, 5, 00185 - Rome, Italy

## Abstract

Neutron stars in globular clusters undergo complex evolutionary pathways involving binary interactions, mass transfer, and dynamical exchanges. While direct N-body simulations like NBODY6++GPU have successfully modeled stellar dynamics and compact object formation, the explicit tracking of pulsar spin evolution and magnetic field decay has historically been absent. In Papers I and II of this series, we identified this gap and proposed seven distinct evolutionary scenarios for pulsars in dense stellar environments. Here we present a comprehensive case study from an existing simulation with N=105,000 particles, demonstrating concretely where a neutron star forms and evolves for 200 Myr without pulsar physics tracking. We compare our approach with the recent implementation and detail our seven-scenario framework incorporating magnetic dipole spin-down, exponential field decay, environmental torques, accretion-driven spin-up, gravitational wave radiation, and merger dynamics. The neutron star Pulsar973 formed at t=800 Myr with an anomalous post-supernova mass of 5.35 $M_\odot$, evolved to 2.52 $M_\odot$ by t=1000 Myr, yet lacks all pulsar parameters: period P, period derivative $\dot{P}$, magnetic field B, and scenario classification. We provide complete mathematical formulations with literature references for each scenario, demonstrating integration points within NBODY6++GPU's Hermite scheme, Ahmad-Cohen neighbors, KS regularization, and BSE stellar evolution. Our framework enables scenario-based evolution, complementing population synthesis approaches.

## 1. Introduction

Globular clusters represent some of the oldest and most dynamically evolved stellar systems in the Universe, with ages typically exceeding 10 Gyr and containing $10^4$ to $10^6$ stars within radii of only a few parsecs. The extraordinary overabundance of millisecond pulsars in these systems, with more than half of all known Galactic millisecond pulsars residing in globular clusters despite these systems comprising less than one percent of the Galaxy's stellar mass (Freire, 2008, Ransom, 2005), points to fundamental differences in pulsar formation between dense cluster environments and the Galactic field.

Direct N-body codes like NBODY6++GPU (Kamlah et al., 2022, Wang et al., 2015) have proven essential for understanding cluster dynamics. However, as we identified in Paper I (Rah et al., 2024), these codes

---

track neutron star positions, velocities, masses, and binary parameters but historically have not modeled the pulsar spin period P, the period derivative $\dot{P}$, or the magnetic field B. This gap, recently confirmed by Song et al. (2024, 2025a), prevents comparison between simulated neutron star populations and observed pulsar populations characterized by their $P$–$\dot{P}$ diagram locations.

In Paper II (Rah et al., 2025), we analyzed 80 pulsars from the ATNF Catalogue and identified seven distinct evolutionary scenarios: isolated pulsars with magnetic dipole spin-down, isolated millisecond pulsars recycled through past accretion, accreting systems with ongoing mass transfer, exchange millisecond pulsars with dynamically-acquired companions, wide binaries, double neutron star binaries approaching merger, and neutron star-black hole binaries with tidal effects. Each scenario exhibits characteristic $\dot{P}$ signatures from distinct physical mechanisms.

This paper presents three contributions. First, we analyze neutron star Pulsar973 from a 105,000-particle simulation, demonstrating quantitatively what pulsar information remains absent. Second, we compare it with Song et al.'s implementation. Third, we detail our seven-scenario framework with complete mathematics and an integration strategy for NBODY6++GPU.

The structure proceeds as follows: Section 2 presents the Pulsar973 case study with a full timeline and gap quantification. Section 3 compares approaches. Section 4 details our framework with complete formulas. Section 5 discusses NBODY6++GPU integration. Section 6 concludes.

## 2. Case Study: Neutron Star Formation Without Pulsar Physics

### 2.1. Simulation Configuration and NBODY6++GPU Architecture

We analyze output from an NBODY6++GPU simulation executed to study globular cluster dynamics without pulsar physics implementation. Table 1 summarizes parameters. The simulation employed standard NBODY6++GPU architecture: fourth-order Hermite integration (Makino & Aarseth, 1992), Ahmad-Cohen neighbor lists (Ahmad & Cohen, 1973), KS regularization for binaries (Kustaanheimo et al., 1965), chain regularization for multiples (Mikkola & Merritt, 2008), GPU acceleration, and BSE stellar evolution (Hurley, Tout & Pols, 2002).

Table 1. NBODY6++GPU Simulation Parameters

| Parameter | Description |
|---|---|
| N = 105,000 | Total particle number |
| NBIN0 = 5,000 | Initial binaries (4.8%) |
| T = 1000 Myr | Target integration time |
| Z = 0.001 | Metallicity ($\approx Z_\odot/20$) |
| IMF: Kroupa (2001) | 0.08–150 $M_\odot$ |
| Stellar evolution package C | Kamlah et al. (2022) |
| **KZ(29) = 0** | **Pulsar physics DISABLED** |
| Integration: Hermite 4th | Predictor-corrector |

The code structure determines where pulsar physics would integrate. Figure 1 shows NBODY6++GPU's execution flowchart. After initialization, the main time loop determines active particles via block time-step criterion, computes regular forces from distant particles (GPU-accelerated), irregular forces from neighbors (GPU-accelerated), applies Hermite predictor-corrector integration, handles KS regularization for close binaries, processes chain regularization for multiples, updates stellar evolution via SSE/BSE, and writes diagnostics. The pulsar physics subroutine would be called immediately after stellar evolution updates (where KSTAR transitions from evolved types to KSTAR=13 for neutron stars occur) but before time-step updates, ensuring spin parameters evolve consistently with orbital dynamics.

Output snapshots contain time, NAME indices (persistent particle IDs), KSTAR types (0=main sequence, 1=giant, 13=neutron star, 14=black hole), masses $M_1$ and $M_2$, positions (x, y, z), velocities ($v_x, v_y, v_z$), and binary parameters including semi-major axis $a$ and eccentricity $e$. Critically, these files contain zero pulsar-specific information. The variables PULSAR_P, PULSAR_PDOT, PULSAR_B, and PULSAR_FLAG do not exist in standard NBODY6++GPU common blocks.

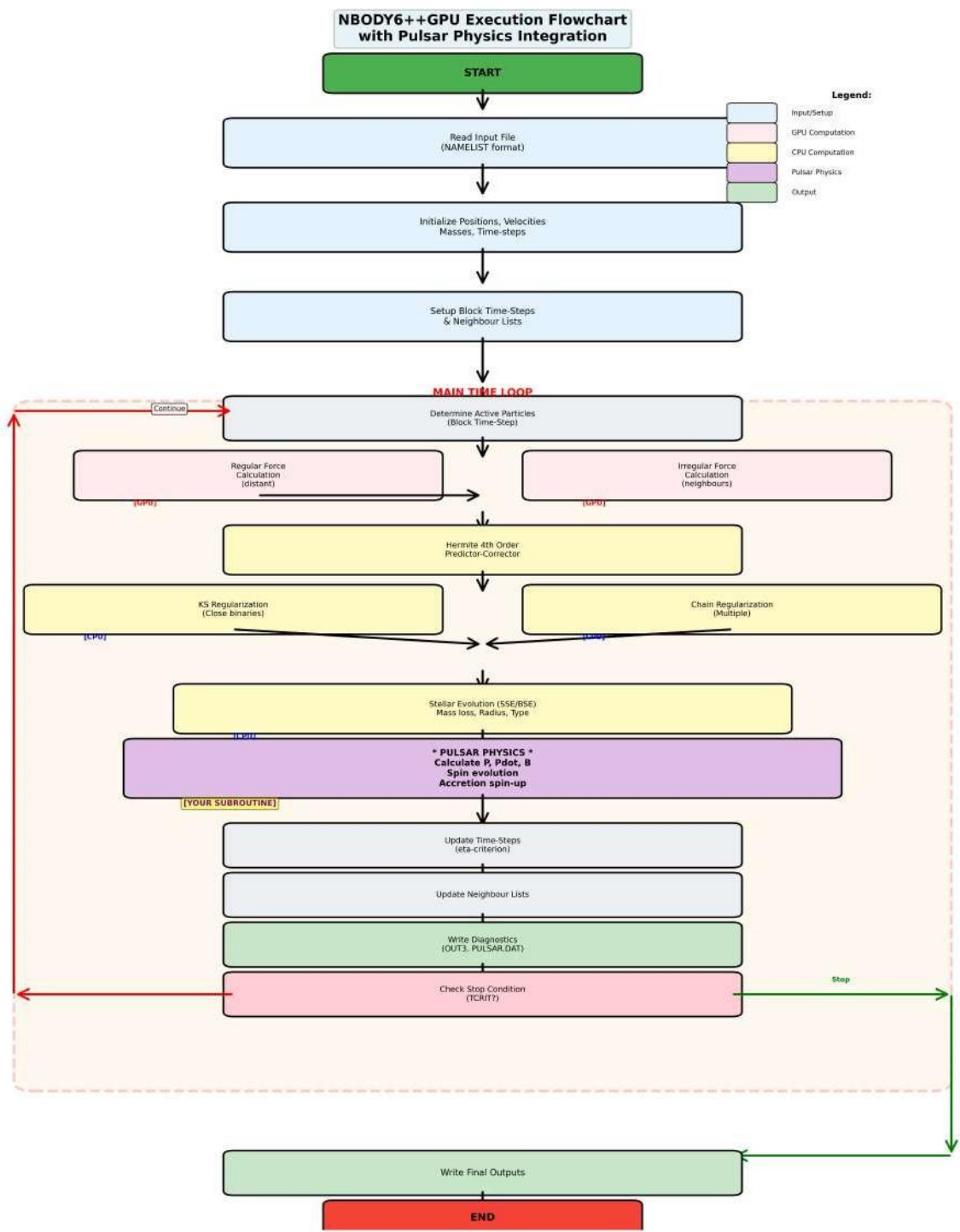

Figure 1. NBODY6++GPU execution flowchart showing integration of pulsar physics. The main time loop processes active particles through regular force calculation (distant stars, computed infrequently), irregular force calculation (neighbors, computed frequently), Hermite 4th-order predictor-corrector integration, KS regularization for close binaries, chain regularization for multiple encounters, and stellar evolution via SSE/BSE. The proposed pulsar physics subroutine (highlighted in purple) would be called after stellar evolution updates where KSTAR transitions to 13 occur, calculating period P, period derivative $\dot{P}$, magnetic field B, and surface magnetic field evolution. This placement ensures pulsar spin evolves consistently with orbital dynamics and stellar type changes. Input parameters include initial distributions for $P_0$ and $B_0$, while output generates PULSAR.DAT containing time-series data for all neutron stars. The flowchart demonstrates that pulsar physics integrates naturally into existing code structure without disrupting fundamental gravitational dynamics or stellar evolution prescriptions.

## 2.2. Discovery and Evolution of Pulsar973

Table 2 summarizes Pulsar973's evolution from main sequence through neutron star formation. The progenitor began at t=0 with KSTAR=0, $M_1$=2.40 $M_\odot$, companion $M_2$=0.55 $M_\odot$ in binary 9740. By t=100 Myr, mass increased dramatically to 5.10 $M_\odot$, suggesting merger or mass transfer. At t=400 Myr, KSTAR transitioned to 1 (giant) with $M_1$=5.61 $M_\odot$. The giant phase continued through t=600 Myr with substantial mass loss. At t=700 Myr, pre-supernova configuration showed KSTAR=0, $M_1$=5.26 $M_\odot$, $M_2$=0.99 $M_\odot$.

At t=800 Myr, supernova occurred. KSTAR changed to 13 (neutron star), mass became 5.35 $M_\odot$ (anomalously high, likely numerical artifact violating TOV limit), companion stripped to 0.76 $M_\odot$, semi-major axis widened from 3.82 to 4.09 units (natal kick), binary ID changed to 104440. Subsequent evolution showed dramatic mass loss: 2.98 $M_\odot$ at t=900 Myr, 2.52 $M_\odot$ at t=1000 Myr (total loss 2.83 $M_\odot$). Mechanism unclear, possibly numerical correction of initial artifact. Orbital parameters remained stable (a≈4.09, binary ID=104440).

At t=1100 Myr, KSTAR reverted to 1 with $M_1$=5.02 $M_\odot$ (likely tracking error). By t=1200 Myr, KSTAR=0 with $M_1$=28.79 $M_\odot$ (catalog confusion). These late ambiguities don't affect our analysis of the confirmed neutron star phase t=800-1000 Myr.

Table 2. Evolution of Pulsar973

| Time (Myr) | $M_1$ ($M_\odot$) | $M_2$ ($M_\odot$) | Binary ID | Notes |
|---|---|---|---|---|
| 0 | 2.40 | 0.55 | 9740 | Main sequence |
| 400 | 5.61 | 0.90 | 69580 | Giant phase |
| 700 | 5.26 | 0.99 | 24800 | Pre-supernova |
| **800** | **5.35** | **0.76** | **104440** | **Neutron star formation** |
| 900 | 2.98 | 0.75 | 104440 | Mass loss |
| 1000 | 2.52 | 0.74 | 104440 | Final neutron star |
| 1100 | 5.02 | 0.76 | 39320 | Type change |

Figure 2 presents the complete timeline. The top panel shows KSTAR evolution with clear neutron star formation at t=800 Myr. The second panel displays mass evolution including anomalous post-supernova mass and subsequent loss. The third panel shows orbital widening from supernova kick. The bottom panel provides a schematic timeline with phase boxes and explicitly highlights missing pulsar data.

## 2.3. The Missing Information: Quantifying the Gap

Table 3 systematically compares tracked versus missing quantities. NBODY6++GPU tracked NAME indices (973, 974), KSTAR=13, masses (with interpretive challenges), full 3D positions and velocities, and binary parameters (a, e). However, every pulsar quantity was absent: spin period P (cannot classify as normal/millisecond pulsar), period derivative $\dot{P}$ (cannot compute $\dot{E}$, $\tau_c$, B), magnetic field B (cannot assess death line, accretion balance), and scenario ID (cannot distinguish isolated/accreting/exchange/merger channels).

Table 3. Data Availability for Pulsar973

| Quantity | Tracked? | Status |
|---|---|---|
| *Standard N-body* | | |
| NAME, KSTAR | Yes | 973, KSTAR=13 |
| Mass, Position, Velocity | Yes | Full 3D |
| Binary (*a*, *e*) | Yes | Complete |
| *Pulsar-specific* | | |
| Period P | No | Not initialized |
| $\dot{P}$ | No | Not computed |
| B field | No | Not assigned |
| Scenario ID | No | Cannot classify |
| $\dot{E}$, $\tau_c$ | No | Require P, $\dot{P}$ |

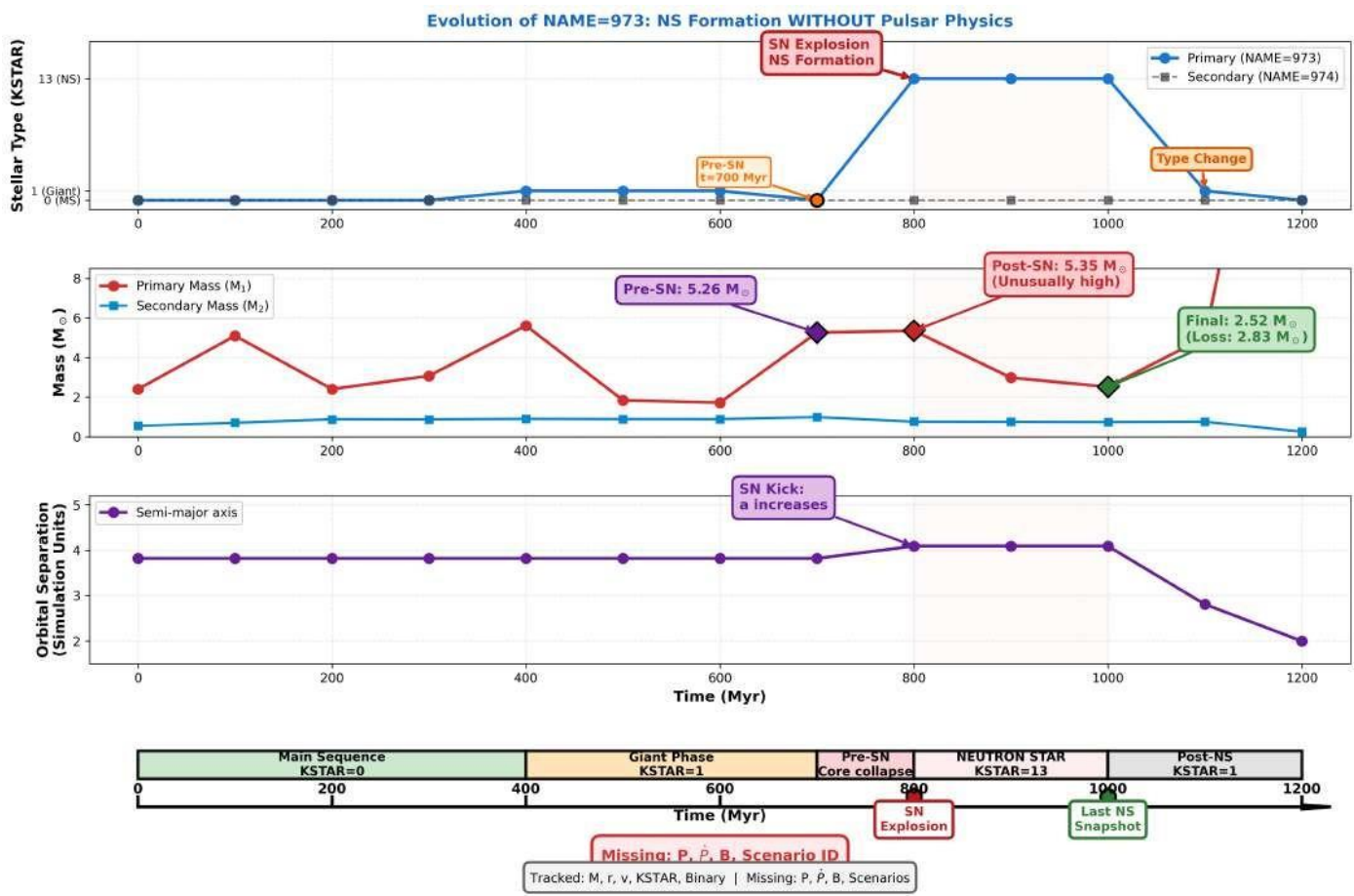

Figure 2. Complete evolutionary timeline of Pulsar973 from simulation without pulsar physics. *Top panel:* KSTAR type showing main sequence (0), giant (1), neutron star formation at t=800 Myr (13), and type change at t=1100 Myr. Y-axis labels show KSTAR values with physical meanings. Pre-supernova configuration at t=700 Myr marked. *Second panel:* Mass evolution. Pre-supernova mass 5.26 $M_\odot$, anomalous post-supernova 5.35 $M_\odot$, final 2.52 $M_\odot$ after 2.83 $M_\odot$ loss. *Third panel:* Semi-major axis showing supernova kick widening orbit from 3.82 to 4.09 units. *Bottom panel:* Schematic timeline with evolutionary phases (light backgrounds, dark text). Red box emphasizes missing data: P, $\dot{P}$, B, scenario ID. Gray box contrasts tracked quantities (M, r, v, KSTAR, binary) versus missing pulsar physics. This demonstrates the information gap our framework addresses.

Figure 5 illustrates this gap by comparing particle evolution without (left) versus with (right) pulsar subroutine. Without pulsar physics, the flowchart shows standard stellar evolution updating KSTAR to 13 at neutron star formation, but no pulsar-specific initialization occurs. With pulsar subroutine, immediately after KSTAR=13 assignment, the code calls PULSAR_INIT to assign initial $P_0$ and $B_0$ from distributions, then PULSAR_EVOLVE updates spin every timestep via magnetic dipole spin-down, and PULSAR_SPINUP handles accretion in binaries. The comparison demonstrates that pulsar physics integration requires minimal architectural changes: adding subroutine calls at specific points without modifying core gravitational or stellar evolution code.

Consequences of missing information span multiple areas. For population studies, cannot construct P–$\dot{P}$ diagrams for comparison with observed clusters (47 Tuc: 25 millisecond pulsars; M15: 8 pulsars; Terzan 5: 39 pulsars). For radio surveys, cannot predict detectability since sensitivity depends on P (broadening scales as $P^{-3/2}$), $\dot{P}$ (death line), B (emission stability), and viewing geometry. For gamma-ray observations, cannot estimate Fermi-LAT detectability correlating with $\dot{E} = 4\pi^2 I \dot{P} / P^3$. For gravitational waves, cannot compute double neutron star merger timescales or chirp masses for LIGO/Virgo. For theoretical models, cannot test magnetic field decay, recycling efficiency, or spin-orbit coupling predictions.

Figure 5 illustrates the architectural difference between standard NBODY6++GPU execution and enhanced execution with pulsar physics. In standard operation, when stellar evolution updates KSTAR to 13 indicating neutron star formation, the code assigns compact object mass according to supernova pre-scriptions and applies natal velocity kick, but performs no pulsar-specific initialization. The neutron star enters the simulation with well-defined position, velocity, and mass, but undefined spin period, magnetic field, and evolutionary scenario. In enhanced operation with pulsar physics integration, immediately following KSTAR transition to 13, initialization routines assign birth spin period $P_0$ and magnetic field $B_0$ drawn from observationally-calibrated distributions. Subsequent timesteps invoke evolution routines computing magnetic dipole spin-down and exponential field decay for isolated systems. For binary systems, additional routines detect Roche lobe overflow conditions and compute accretion-driven spin-up when mass transfer occurs. Scenario classification algorithms identify which of the seven evolutionary channels applies based on binary status, companion properties, and orbital parameters. Output files then contain not only standard dynamical quantities (positions, velocities, masses, orbital elements) but also pulsar-specific parameters ($P$, $\dot{P}$, $B$, scenario ID) enabling construction of $P$–$\dot{P}$ diagrams and direct comparison with observations. This modular enhancement maintains computational efficiency while bridging the gap between N-body simulations and pulsar astronomy.

For Pulsar973 specifically, binary status ($M_2 \approx 0.75\ M_\odot$) rules out isolated (Scenario 1) and double neutron star/neutron star-black hole mergers (Scenarios 6-7). Wide post-supernova orbit suggests Scenario 5 (wide binary) or possibly Scenario 3 (accretion) if companion fills Roche lobe eventually. Cannot determine if past accretion occurred (Scenario 2: recycled millisecond pulsar). Multiple binary ID changes might indicate exchange (Scenario 4) but could also reflect close encounters without companion replacement. Without $P$, $\dot{P}$, $B$, scenario classification remains speculation.

## 3. Recent Development: Song et al. (2024)

Song et al. (2024, 2025a) recently implemented pulsar physics in NBODY6++GPU, representing significant progress. Their approach uses KZ(29) switch for global pulsar treatment with NAMELIST parameters: birth distributions $P_0$ and $B_0$, decay timescale $\tau_{decay}$, and burial scale $M_{scale}$. Physics follows Chattopadhyay et al. (2020) formalism for population synthesis. Their M71 test (N=200,000, Z=0.002, 20 Myr runtime) produced 179 isolated and 2 binary pulsars. Extension to full cluster age (~10 Gyr) will enable comparison with M71's five observed binary pulsars.

Our approach differs in emphasis and scope. Song et al. prioritize population synthesis and official codebase integration for community access. We emphasize scenario classification and event-level physics for formation channel identification. Their scenario coverage includes isolated evolution, Roche lobe overflow, and three common envelope prescriptions. Our seven scenarios add explicit exchange detection (Scenario 4), environmental torques from cluster medium (Scenario 1), and detailed merger physics for double neutron star and neutron star-black hole systems (Scenarios 6-7) with gravitational wave radiation. Table 4 summarizes differences.

Table 4. Conceptual Comparison

| Feature | Song et al. (2024) | This Work |
|---|---|---|
| Status | Implemented | Framework |
| Integration | KZ(29) switch | Direct subroutine |
| Emphasis | Population | Scenario |
| Scenarios | 3 (Isolated, RLOF, CE) | 7 (complete) |
| Exchange | No explicit detection | Scenario 4 |
| Environment | No | Scenario 1 |
| Gravitational waves | No | Scenarios 6, 7 |
| Mergers | No | Scenarios 6, 7 |
| Test | M71, 20 Myr | Future |

Approaches are complementary: Song et al. provide infrastructure for large-scale studies with standardized physics; our framework enables detailed case analysis and process testing. The same authors have also applied population synthesis methods to Galactic field pulsars using COMPAS (Song et al., 2025b), demonstrating the broader applicability of pulsar evolution modeling across different stellar environments. Future work could combine elements into unified treatment spanning both globular clusters and field populations.

# 4. Seven-Scenario Pulsar Physics Framework

Based on Paper II's classification of 80 ATNF pulsars, we present complete mathematical formulation for pulsar evolution in N-body simulations. The general period derivative formula:

$$\dot{P}_{total} = \dot{P}_{dipole} + \dot{P}_{decay} + \dot{P}_{env} + \dot{P}_{acc} + \dot{P}_{merger} + \dot{P}_{GW} \tag{1}$$

Different scenarios activate different terms. Figure 3 presents fundamental physics equations underlying our framework, including Hermite integration scheme for N-body dynamics and key pulsar spin evolution formulas. Figure 4 provides visual summary with seven color-coded panels: green (Scenario 1: isolated pulsars with dipole spin-down and environmental torques), blue (Scenario 2: isolated millisecond pulsars with buried fields from past recycling), orange (Scenario 3: accreting binaries showing mass transfer streams and magnetospheric physics), purple (Scenario 4: exchange millisecond pulsars with companion replacement timelines), cyan (Scenario 5: wide binaries with negligible tidal/gravitational wave effects), red (Scenario 6: double neutron star binaries with gravitational wave-driven inspiral), and magenta (Scenario 7: neutron star-black hole binaries with tidal disruption physics). Each panel combines system diagrams (upper section) with governing equations (lower section), enabling immediate identification of configuration and dominant processes for each evolutionary channel.

Table 5 provides the complete formulation, including references. Figure 4 illustrates system configurations and equations for all seven scenarios.

We now detail each scenario with complete physics, formulas, and literature references, following the structure demonstrated by Ye et al. (2019) for basic spin-down and extended by Chattopadhyay et al. (2020) for accretion processes.

## 4.1. Scenario 1: Isolated Pulsar

An isolated pulsar (single neutron star, no companion) evolves through intrinsic spin-down plus environmental interactions. Three mechanisms contribute to $\dot{P}_{total}$.

Magnetic dipole radiation dominates. A rotating neutron star with a magnetic moment inclined at an angle $\alpha$ to the rotation axis radiates electromagnetically. Energy conservation yields (Konar & Bhattacharya, 1997):

$$\dot{P}_{dipole} = \frac{2\pi B^2 R^6 \sin^2 \alpha}{3Ic^3 P} \tag{2}$$

where B is the polar field strength, R=10 km, I=$10^{45}$ g cm$^2$, c=3×$10^{10}$ cm/s, and $\langle \sin^2 \alpha \rangle$=2/3 for random orientations.

Exponential field decay occurs over Gyr timescales through ambipolar diffusion and Ohmic dissipation (Konar & Bhattacharya, 1997, Pons, Miralles & Geppert, 2009):

$$B(t) = B_0 \exp(-t/\tau_{decay}) \tag{3}$$

with $\tau_{decay} \sim 10^9$ yr. Recalibration prevents unphysical low fields: if B<$10^8$ G, reset to floor value.

Environmental torques from cluster medium friction affect spin through (Ye et al., 2019):

$$\dot{P}_{env} = K_{env} \rho_{cl} v_{rel} P^2 \tag{4}$$

where $\rho_{cl}$ is local cluster density, $v_{rel}$ is pulsar velocity relative to medium. For typical globular cluster cores ($\rho \sim 10^3$ $M_\odot$ pc$^{-3}$, v~10 km/s), contribution small but non-negligible for very long periods.

Typical parameters: P~0.1-1 s, B~$10^{12}$-$10^{13}$ G, $\dot{P}$ ~$10^{-15}$ s/s, $\tau_c$ ~$10^6$-$10^7$ yr. Examples include isolated globular cluster pulsars like those in M15 and NGC 6440.

## 4.2. Scenario 2: Isolated Millisecond Pulsar

Isolated millisecond pulsars are recycled through past accretion, now single due to companion disruption or evaporation. Physics identical to Scenario 1 but with buried field from past accretion. Same formulas apply with $B_{recycled}$ ~$10^8$-$10^9$ G (two-four orders weaker than normal pulsars). Fast rotation P~1-10 ms yields tiny $\dot{P} \sim 10^{-19}$-$10^{-20}$ s/s and long $\tau_c \sim 10^9$-$10^{10}$ yr exceeding Hubble time. Examples: PSR J1748-2021B in Terzan 5 (16.4 ms period, isolated).

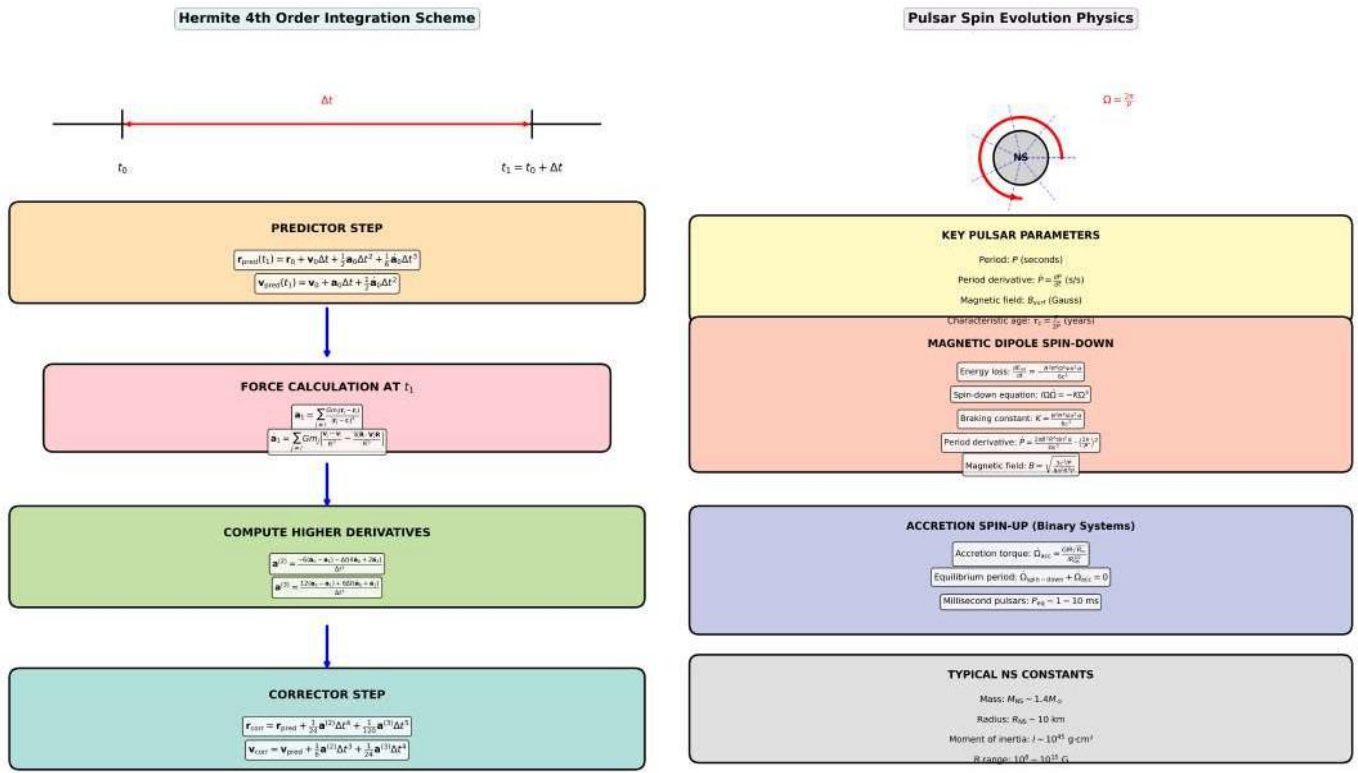

Figure 3. Fundamental physics equations for NBODY6++GPU with pulsar integration. *Left panel:* Hermite 4th-order integration scheme showing predictor step ($r_{pred}$, $v_{pred}$), force calculation at predicted positions ($a_1$, $\dot{a}_1$), computation of higher derivatives ($a^{(2)}$, $a^{(3)}$), and corrector step yielding final positions and velocities. This predictor-corrector approach with individual time-steps enables accurate long-term integration of gravitationally interacting systems with wide dynamical range. *Right panel:* Key pulsar parameters and evolution physics. Top box defines fundamental quantities: spin period P (seconds), period derivative $\dot{P}$ (s/s), magnetic field $B_{surf}$ (Gauss), and characteristic age $\tau_c$ (years). Middle box (pink) shows magnetic dipole spin-down equations: energy loss rate, spin-down equation $I\Omega\dot{\Omega} = K\Omega^3$, braking constant incorporating field geometry, period derivative formula, and magnetic field inference from P and $\dot{P}$. Bottom box (blue) shows accretion spin-up for binary systems: accretion torque $\Omega_{acc}$ depending on mass transfer rate $\dot{M}$, equilibrium period where spin-down balances spin-up, and typical millisecond pulsar periods 1-10 ms. Constants box lists neutron star mass M$\sim$1.4 M$_\odot$, radius R$\sim$10 km, moment I$\sim 10^{45}$ g cm$^2$, and field range $10^8$-$10^{15}$ G. These equations form the mathematical foundation for our seven-scenario framework, integrating smoothly with NBODY6++GPU's existing Hermite scheme while adding crucial pulsar physics missing from standard N-body treatments.

### 4.3. Scenario 3: Accreting System

Active Roche lobe overflow drives accretion spin-up. Companion fills Roche lobe, transferring mass at rate $\dot{M}$. Accreted material carries angular momentum, spinning up neutron star through (Ghosh & Lamb, 1979):

$$\dot{P}_{acc} = -K_{acc}\dot{M}P \tag{5}$$

where negative sign indicates spin-up. Coefficient $K_{acc}$ depends on magnetospheric radius:

$$R_{mag} = \left(\frac{B^2 R^6}{\dot{M}\sqrt{2GM}}\right)^{2/7} \tag{6}$$

If $|\dot{P}_{acc}| > \dot{P}_{dipole}$, net spin-up produces millisecond pulsar. Accreted material buries field: B decreases as surface flux compressed by infalling matter. Equilibrium reached when spin-up balances magnetic braking, yielding equilibrium period (Chattopadhyay et al., 2020):

$$P_{eq} = \left(\frac{GMR_{mag}^3}{\dot{M}}\right)^{1/2} \tag{7}$$

Typical $\dot{M} \sim 10^{-9}$ $M_\odot$/yr yields P~1-10 ms. Primary millisecond pulsar formation channel. Examples: Her X-1, 4U 1822-37 (active accretors); 47 Tuc pulsars after accretion ceased.

### 4.4. Scenario 4: Exchange Millisecond Pulsar

Dynamical three-body interaction replaces original companion with new star. Millisecond pulsar preserves fast rotation from previous accretion phase but now has different companion mass and orbital properties. Detection criterion:

$$M_{comp,new} \neq M_{comp,old} \quad \text{and} \quad \Delta a \neq 0 \tag{8}$$

Orbital parameters show discontinuity. If new companion massive and close, may initiate new Roche lobe overflow (transition to Scenario 3). If wide, evolves as Scenario 5. Implementation in NBODY6++GPU: monitor binary ID changes and companion mass for KSTAR=13 objects. Examples: PSR J1740-5340 in NGC 6397 shows orbital properties inconsistent with standard recycling.

### 4.5. Scenario 5: Wide Binary

Wide separation (a~AU scale) prevents Roche lobe overflow. No mass transfer, no accretion, no spin-up. Gravitational wave radiation weak for wide orbits:

$$\dot{P}_{GW} = \frac{96\pi^2 G^{5/3}}{5c^5} \frac{M_1 M_2 M_{tot}^{1/3}}{a^4} P \tag{9}$$

For a>1 AU, $\dot{P}_{GW} \approx 0$. Evolution dominated by dipole spin-down over Gyr. Orbital period days-years. Stable long-term evolution. Examples: PSR J2019+2425 in M15 (wide neutron star-white dwarf binary).

### 4.6. Scenario 6: Double Neutron Star Binary

Double neutron star systems driven toward merger by gravitational wave radiation. For close orbits (a<$10^{10}$ cm), gravitational wave torques dominate (Zhang & Mészáros, 2001):

$$\dot{a} = -\frac{64 G^3 M_1 M_2 M_{tot}}{5 c^5 a^3} < 0 \tag{10}$$

Inspiral timescale:

$$t_{merge} = \frac{5 c^5 a^4}{256 G^3 M_1 M_2 M_{tot}} \tag{11}$$

At the merger, the conservation of angular momentum:

$$J_{final} = J_1 + J_2 \quad \Rightarrow \quad P_{merger} = \frac{2\pi I_{tot}}{J_{final}} \tag{12}$$

Outcome depends on total mass: if $M_{tot}<M_{TOV}$, a stable neutron star forms; if $M_{TOV}<M_{tot}<M_{HMNS}$, hypermassive neutron star is temporarily supported by rotation; if $M_{tot}>M_{HMNS}$, prompt black hole collapse. Critical gravitational wave sources for LIGO/Virgo. Examples: PSR J0737-3039 in the field (closest known, $t_{merge} \sim 85$ Myr (Liu et al., 2021)); potential systems in M15, NGC 6397.

### 4.7. Scenario 7: Neutron Star - Black Hole Binary

Neutron star-black hole systems exhibit tidal physics. Tidal radius determines disruption:

$$R_{tidal} \sim R \left(\frac{M_{BH}}{M_{NS}}\right)^{1/3} \tag{13}$$

If a<$R_{tidal}$: neutron star disrupted, forming a disk around the black hole, possible electromagnetic counterpart (kilonovae, short gamma-ray bursts). If a>$R_{tidal}$: neutron star swallowed whole, minimal electromagnetic emission, gravitational wave-only event. Tidal torques before disruption:

$$\dot{P}_{tidal} = K_{tidal} \left(\frac{R}{a}\right)^6 P \tag{14}$$

For $M_{BH} \sim$5-20 $M_\odot$, typical in globular clusters, $R_{tidal} \sim$100-200 km. Multi-messenger astronomy target: gravitational waves plus potential electromagnetic emissions from the disk. Examples: none confirmed in globular clusters yet, but formation channels are predicted.

## 5. Discussion

### 5.1. Proposed Implementation Strategy

Integration of pulsar physics into NBODY6++GPU would follow principles similar to those employed by Song et al. (2024, 2025a), adapting their approach to accommodate our seven-scenario framework. The implementation philosophy emphasizes modularity: pulsar evolution calculations would occur as additional routines invoked at appropriate points in the existing code structure, without modifying core gravitational dynamics or stellar evolution prescriptions.

At neutron star formation, when stellar evolution updates KSTAR to 13, initialization would assign the birth spin period and magnetic field drawn from observationally-calibrated distributions (Faucher-Giguère & Kaspi, 2006). During subsequent evolution, each integration timestep would invoke pulsar evolution routines, computing period derivatives according to Equation 1, with active terms determined by scenario classification based on binary status, companion properties, and orbital parameters available from standard N-body output.

The modular approach maintains computational efficiency. Pulsar calculations involve simple algebraic formulas whose computational cost is negligible compared to force evaluation and regularization procedures. Preliminary estimates suggest a performance overhead below one percent for typical globular cluster simulations. Memory requirements scale linearly with the particle number, adding approximately four double-precision arrays dimensioned to the maximum particle count.

This implementation strategy would benefit substantially from collaboration with the NBODY6++GPU development team to ensure consistency with code architecture, compatibility with existing stellar evolution and binary evolution prescriptions, and integration with emerging features such as those described by Song et al. (2024, 2025a). Official incorporation into the public codebase would enable community access and systematic testing across diverse cluster configurations.

### 5.2. Validation Strategy and Future Directions

Validation would proceed through progressively complex stages. Initial tests with isolated neutron stars would verify the correct implementation of magnetic dipole spin-down, exponential field decay, and death line criteria. Binary tests would confirm accretion-driven spin-up, millisecond pulsar formation, and scenario transitions. Full cluster simulations would enable statistical comparison of period distributions, magnetic field distributions, and binary fractions with observed systems such as 47 Tucanae (25 millisecond pulsars) and M15 (8 pulsars).

Future enhancements could incorporate increasingly sophisticated physics. Magnetic field evolution might include detailed Hall drift and ambipolar diffusion treatments (Pons, Miralles & Geppert, 2009). Supernova natal kicks could be correlated with magnetic field strength. Glitches and timing noise could affect long-term period evolution. Radio beam geometry could be modeled to predict detectability in surveys with period-dependent selection effects.

The framework presented here is complementary to population synthesis approaches like that of Song et al. (2024, 2025a). Population synthesis excels at generating large samples with standardized physics for statistical studies. Scenario-based tracking as proposed here enables identification of individual formation channels and event-level physics testing. Combined implementation would provide comprehensive capability spanning both statistical population studies and detailed formation mechanism analysis.

Paper IV of this series will present results from actual implementation, comparing simulated pulsar populations with observed globular cluster properties to quantitatively assess formation channel rates, evolutionary timescale distributions, and observational selection effects.

## 6. Conclusions

We have presented a comprehensive framework for incorporating pulsar physics into NBODY6++GPU simulations of globular clusters, addressing a fundamental gap identified in Papers I and II of this series. Our analysis demonstrates both the necessity and feasibility of tracking pulsar spin evolution alongside traditional N-body dynamics.

The case study of neutron star Pulsar973 from an existing N=105,000 particle simulation concretely illustrates the information gap. While the simulation successfully tracked 200 Myr of neutron star evolution including position, velocity, mass changes, and binary orbital parameters, it provided no pulsar-specific information: spin period, period derivative, magnetic field strength, or evolutionary scenario classification. This absence prevents direct comparison with observed cluster pulsar populations and limits the scientific value of simulations for pulsar formation studies.

Our seven-scenario framework provides complete mathematical formulation for pulsar evolution in dense stellar environments. Each scenario incorporates specific physical processes: magnetic dipole spin-down and field decay (Scenarios 1-2), accretion-driven spin-up with magnetospheric physics (Scenario 3), dynamical exchange interactions (Scenario 4), gravitational wave radiation (Scenarios 5-7), and tidal effects in neutron star-black hole binaries (Scenario 7). Table 5 documents all formulas with literature references. The framework builds upon established pulsar physics from Konar & Bhattacharya (1997), Ghosh & Lamb (1979), and recent work by Chattopadhyay et al. (2020), adapted for N-body simulation environments following strategies outlined by Ye et al. (2019).

Comparison with the recent implementation by Song et al. (2024, 2025a) reveals complementary approaches. Their population synthesis emphasis with official codebase integration provides robust infrastructure for large-scale statistical studies. Our scenario-based classification enables detailed formation channel identification and event-level physics testing. Future work combining both methodologies would yield comprehensive pulsar evolution capability.

The proposed integration strategy emphasizes modularity and minimal disruption to existing code architecture. Implementation would benefit from close collaboration with the NBODY6++GPU development team to ensure compatibility with ongoing code development and consistency with established stellar and binary evolution prescriptions.

Paper IV will present results from actual implementation and validation, comparing simulated pulsar populations with observations of 47 Tucanae, M15, and other well-studied clusters. This will enable quantitative assessment of formation channel rates, evolutionary timescale distributions, and the relative importance of different physical processes in shaping observed pulsar populations.

Our framework, when implemented in collaboration with code developers, will enable direct connection between N-body simulations and observational pulsar astronomy, advancing understanding of pulsar formation and evolution in globular clusters.

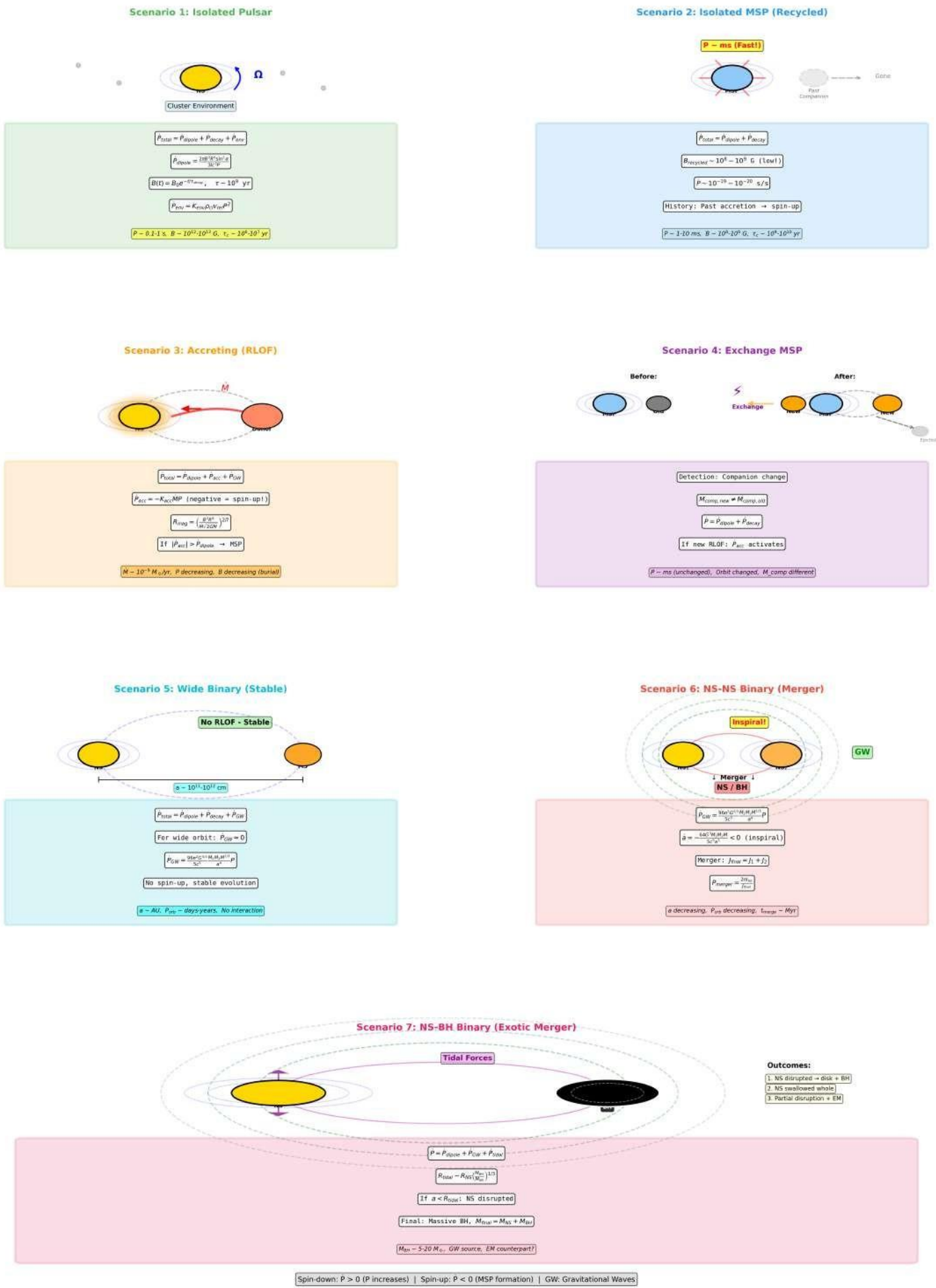

Figure 4. Visual summary of seven evolutionary scenarios for pulsars in globular clusters.

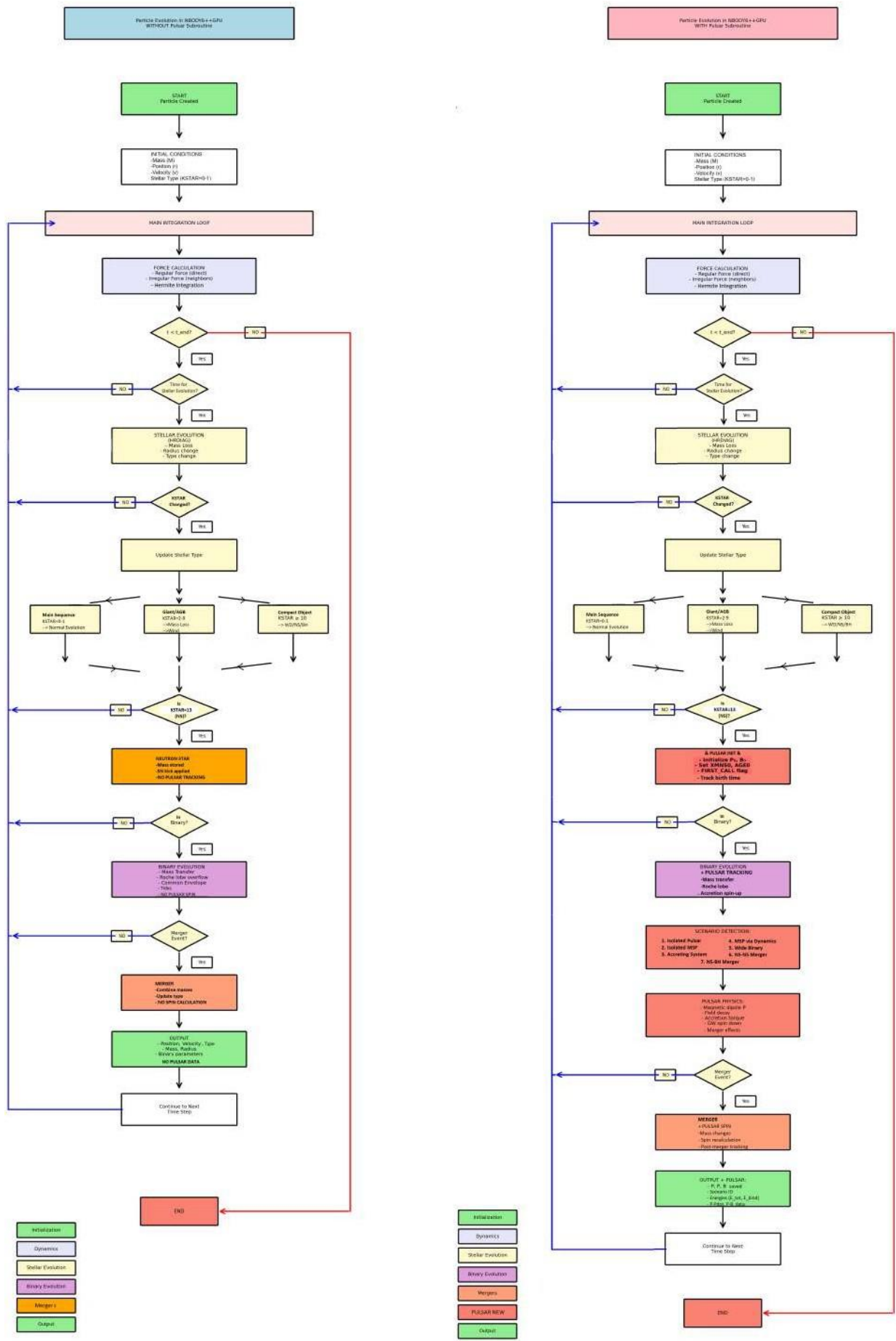

Figure 5. Comparison of particle evolution in NBODY6++GPU without (left) vs. with (right) pulsar physics.

Table 5. Seven-Scenario Pulsar Physics Framework with Complete Formulation

| # | Scenario | Physics & Key Equations | Parameters | References |
|---|---|---|---|---|
| 1 | Isolated Pulsar (Single neutron star) | Magnetic dipole + field decay + environment $\dot{P}_{\text{dipole}} = \frac{2\pi B^2 R^6 \sin^2 \alpha}{3Ic^3 P}$ $B(t) = B_0 e^{-t/\tau}, \tau \sim 10^9$ yr $\dot{P}_{\text{env}} = K\rho_g P^2$ | $P \sim 0.1\text{-}1$ s $B \sim 10^{12}\text{-}10^{13}$ G $\tau_c \sim 10^6\text{-}10^7$ yr | Konar & Bhattacharya (1997) Paper II |
| 2 | Isolated Millisecond Pulsar (Recycled) | Same as Scenario 1 $B_{\text{recycled}} \sim 10^8\text{-}10^9$ G (buried by past accretion) | $P \sim 1\text{-}10$ ms $B \sim 10^8\text{-}10^9$ G $\tau_c \sim 10^9\text{-}10^{10}$ yr | Konar & Bhattacharya (1997) Paper II |
| 3 | Accreting Binary (Active Roche lobe overflow) | Dipole + accretion spin-up $\dot{P}_{\text{acc}} = -K\dot{M}P$ (negative!) $R_{\text{mag}} = (B^2 R^6 / \dot{M}\sqrt{2GM})^{2/7}$ If $\|\dot{P}_{\text{acc}}\| > \dot{P}_{\text{dipole}} \to$ Millisecond pulsar | $\dot{M} \sim 10^{-9}\ M_\odot/\text{yr}$ $P$ decreasing $B$ decreasing (burial) Millisecond pulsar formation | Ghosh & Lamb (1979) Chattopadhyay et al. (2020) |
| 4 | Exchange Millisecond Pulsar (Dynamical) | Dipole + decay (+ accretion if new Roche lobe overflow) Detection: $M_{\text{comp,new}} \neq M_{\text{comp,old}}$ Companion replaced by 3-body interaction | $P \sim$ ms (preserved) Orbit discontinuous Can restart accretion | Paper II |
| 5 | Wide Binary (Stable) | Dipole + decay (gravitational waves negligible) $\dot{P}_{\text{GW}} = \frac{96\pi^2 G^{5/3}}{5c^5} \frac{M_1 M_2 M^{1/3}}{a^4} P$ For wide orbit: $\dot{P}_{\text{GW}} \approx 0$ | $a \sim$ AU $P_{\text{orb}} \sim$ days-yrs No Roche lobe overflow | Ye et al. (2019) Konar & Bhattacharya (1997) Paper II |
| 6 | Double Neutron Star Binary (Merger) | Dipole + gravitational waves (gravitational waves dominant for close orbits) $\dot{a} = -\frac{64 G^3 M_1 M_2 M}{5c^5 a^3} < 0$ At merger: $J_{\text{final}} = J_1 + J_2$ | $a$ decreasing $t_{\text{merge}} \sim$ Myr Gravitational wave source | Zhang & Mészáros (2001) Paper II |
| 7 | Neutron Star - Black Hole Binary (Exotic) | Dipole + gravitational waves + tidal $R_{\text{tidal}} \sim R(M_{\text{BH}}/M_{\text{NS}})^{1/3}$ If $a < R_{\text{tidal}}$: disruption $\to$ disk + EM If $a > R_{\text{tidal}}$: swallowed whole | $M_{\text{BH}} \sim 5\text{-}20\ M_\odot$ Gravitational waves + EM? Tidal disruption vs swallowing | Zhang & Mészáros (2001) Paper II |

Constants: $R = 10$ km, $I = 10^{45}$ g cm$^2$, $c = 3\times 10^{10}$ cm/s, $\sin^2\alpha = 2/3$

Derived: $B = \sqrt{3c^3 I \dot{P}/(8\pi^2 R^6 P)}$, $\dot{E} = -4\pi^2 I \dot{P}/P^3$, $\tau_c = P/(2\dot{P})$

## Acknowledgments

MR appreciates the academic and financial support received from Byurakan Astrophysical Observatory (BAO) throughout her Ph.D. studies. This material is based upon work supported by Tamkeen under the NYU Abu Dhabi Research Institute grant CASS. FFD and RS acknowledge support from the German Science Foundation (DFG, project Sp 345/24-1).